\documentclass[pra,twocolumn,showpacs,superscriptaddress]{revtex4-1}  % default: %\bibliographystyle{apsrev4.bst}
\usepackage{natbib}
\usepackage{graphicx}
\usepackage{subfigure}
\usepackage{epsfig} % epsf no longer supported in 4.1
\usepackage{amsmath}
\usepackage{color}
\usepackage{multirow}
\begin{document}

\title{Photon Hall Scattering from Alkaline-earth-like atoms and Alkali-like ions} 
%\\ A Gedanken Experiment in Quantum Physics ?}

\author{B.A. van Tiggelen}
\affiliation{Universit\'{e} Grenoble-Alpes, Laboratoire de Physique et de Mod\'elisation des
Milieux Condens\'es, Grenoble, France}
\affiliation{CNRS, Laboratoire de Physique et de Mod\'elisation des
Milieux Condens\'es, Grenoble, France}
%\author{M. Mukhtar}
%\altaffiliation[Permanent address: ]{Laboratoire Charles-Fabry, Institut d'Optique, CNRS, Universit\'{e} Paris-Sud, Universit\'{e} %Paris-Saclay,  Orsay, France.}
%\affiliation{Centre for Quantum Technologies, National University of Singapore, 117543 Singapore}
\author{D. Wilkowski}
\affiliation{MajuLab, CNRS-Universit\'{e} de Nice-NUS-NTU International Joint Research Unit UMI 3654, Singapore}
\affiliation{Centre for Quantum Technologies, National University of Singapore, 117543 Singapore}
\affiliation{School of Physical and Mathematical Sciences, Nanyang Technological University, 637371, Singapore}

\date{\today}

\begin{abstract}
We investigate the possibility of observing a magneto-transverse scattering of photons from alkaline-earth-like atoms as well as alkali-like ions and provide orders of magnitude. The transverse magneto-scattering is physically induced by the interference between two possible quantum transitions of an outer electron in a $S$ state, one dispersive electric-dipole transition to a $P$ orbital state and a second resonant electric-quadrupole transition to a $D$ orbital state. In contrast with previous mechanisms proposed for such an atomic photonic Hall effect, no real photons are scattered by the electric-dipole allowed transition, which increases the ratio of Hall current to background photons significantly. The main experimental challenge is to overcome the small detection threshold, with only $10^{-5}$ photons scattered per atom per second.
\end{abstract}

%\pacs{42.25.Dd, 43.20.Gp, 71.23.An}
.

\maketitle

\section{Introduction}
Atoms and ions are quantum objects that can scatter photons.  This well-known process involves the promotion of an electron to an excited quantum level at the cost of a photon, and its subsequent decay with the emission of a new photon, neither necessarily in the same direction nor at the same frequency. Because in vacuum, atoms are small compared to the excitation wavelength $\lambda$, atom-photon scattering is strongly restricted by selection rules, with the electric-dipole matrix elements strongly dominating over higher electric-multipoles. In addition, because of spin-orbit coupling and electron-electron interaction, the degeneracy of quantum levels associated with different angular momenta is largely lifted. As a result, the quantum-interference between e.g. a
$S \leftrightarrow P$  and a $S\leftrightarrow D$ transition is usually not relevant.

The Photon Hall Effect (PHE) is induced by the interference between transitions associated with different multipoles, somewhat  similar to rotatory power in chiral molecules \cite{book}. It was first observed 20 years ago in multiple scattering samples \cite{geert}. In classical Mie scattering such interference poses no problem, especially for particles of the size of the wavelength \cite{PHE}, where many multipoles coexist at not too small frequencies. For atoms such interference is rare and small. Two close atoms can generate together an electric quadrupole (EQ) out of their individual electric dipoles (ED) and induce a PHE \cite{Anja,DD}. For one atom, only the hydrogen atom - with almost degenerated levels $^2P_{3/2}$ and $^2D_{3/2}$ - has been proposed recently \cite{PHEatom}. Atomic hydrogen however, is difficult to handle experimentally. Its PHE is also complicated by the giant resonant absorption of the ED transition that accelerates the atom and makes it move out of resonance quickly. Nevertheless, if these technical problems would be solved, a PHE would be generated, equal to a magneto-transverse acceleration of order $3$\,m/s$^2$ for a broadband incident flux of $10$\,kW/m$^2$ and magnetic fields of order of $10$ Gauss \cite{PHEatom}.

\begin{figure}
   \begin{center}
      \includegraphics[width =0.7\linewidth]{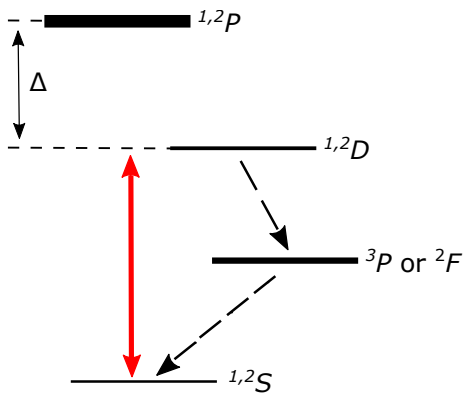}
      \caption{Energies and transitions of interest for alkali-like neutral atom or ion. The PHE effect originates from the interference of the $S\leftrightarrow P$ ED transition with the $S\leftrightarrow D$ EQ transition. The excitation laser is tuned on the EQ resonance (Red arrow). Beside $D\rightarrow S$ elastic fluorescence, possible other decay channels towards intermediate spin triplet $^3P$ states for neutral atoms or spin doublet $^2F$ states for ions are indicated. Those states are supposed to decay rapidly towards the ground state, either with or without the help of re-pumping lasers beams.}
      \label{Fig1}
   \end{center}
\end{figure}

 In this paper we explore the PHE effect for neutral atoms and ions where laser cooling and trapping technics are currently available. Those elements should have in common an energy level structure sketched in Fig.~\ref{Fig1}. One needs a $S\leftrightarrow P$ ED transition and a $S\leftrightarrow D$ EQ transition. This general property is encountered in many elements. However, since we are dealing with elastic scattering of photons on resonance with the EQ transition, it is important that the decay of the excited $D$ state will not dominated by other inelastic fluorescence channels. This requirement excludes the commonly used alkali atoms, where a strong $D\rightarrow P$ intermediate ED transition exists. As example the $5\,^2D_{5/2}\rightarrow 6\,^2S_{1/2}$ transition of Cs is counting only for $3\times 10^{-5}$ of the total decay rate of the $5\,^2D_{5/2}$ state \cite{chan2016doppler}. Recently cooled rare-earth atoms such as Er \cite{PhysRevLett.96.143005}, Dy \cite{PhysRevLett.104.063001}, Tm \cite{PhysRevA.82.011405} and also Cd \cite{PhysRevA.76.043411} and Hg \cite{hachisu2008trapping}, should be excluded for similar reasons. Moreover, scattering of photons on the EQ transition require a relatively large EQ coupling. This second requirement excludes several ions currently used for optical clocks like, for example, Ca$^+$ \cite{PhysRevA.77.021401}. We propose four promising elements, summarized in Table \ref{table}, the two neutral alkaline-earth atoms $^{40}$Ca, $^{88}$Sr, and the two alkali-like ions $^{88}$Sr$^+$ and $^{174}$Yb$^+$ \cite{0022-3700-18-8-011,PhysRevA.56.2699}. For those elements, the energy difference between the $P$ and the $D$ states is more than one million times larger than the natural linewidth of the ED transition. Thus, even if the ED transition is much stronger than the EQ transition, for a laser tuned on the EQ resonance, the population of the $P$ state is negligible, and the optical absorption by the ED transition can be disregarded. In section \ref{PHS}, we show how the interference of the EQ transition with the dispersive response of the ED transition creates a PHE, if a bias DC magnetic-field is applied to split the Zeeman substates of the $D$-level. In section \ref{Implementation}, we discuss the practical implementation of a PHE experiment considering in particular the unavoidable Doppler effect on the EQ transition, the inelastic scattering, and the radiation pressure force.

% Figure 1 %%%%%%%%%%%%%%%%%%%%%%%
%%%%%%%%%%%%%%%%%%%%%%%%%%%%%%%%%%

\section{Photon Hall Scattering}
\label{PHS}
\begin{table}[h]
\begin{tabular}{| c | c | c | c | c |}
                \hline
               Element & Sr & Ca & Yb$^+$ & Sr$^+$ \\
               \hline\hline
               ED line & \multicolumn{2}{|c|}{$^1S_0\leftrightarrow\,^1P_1$} & \multicolumn{2}{|c|}{$^2S_{1/2}\leftrightarrow\,^2P_{3/2}$} \\
               \hline
              $\lambda_P$ (nm) & 461  & 423 & 329  & 408 \\
              \hline
              $A_P/2\pi$ (MHz) & 32 & 34 & 19.6 & 22 \\
              \hline\hline
              EQ line & \multicolumn{2}{|c|}{$^1S_0\leftrightarrow\,^1D_2$} & \multicolumn{2}{|c|}{$^2S_{1/2}\leftrightarrow\,^2D_{5/2}$}\\
              \hline
              $\lambda_D$ (nm) & 496 & 458 & 411 & 674 \\
              \hline
              $A_D/2\pi$ (Hz) & 9.5 & 6.2 & 3.8 & 0.4 \\
              \hline
              $A_D/\Gamma_D$ & 0.02  & 0.12 &  0.17 & 1\\
              \hline
              $I_s$ (mW/m$^2$) &  770 & 18 & 7.5 & 0.005\\
              \hline\hline
              $A_P/\Delta_{PD} \times 10^7$ & 6.9 & 6.3 & 1.1 & 0.76\\
              \hline \hline
              $\sigma_{\mathrm{PHE}}$ (pm$^2$) &  80.4 & 372 & 48 & 220 \\
              \hline
              $2\sigma_{\mathrm{PHE}}/\sigma_{\mathrm{EQ}} \times 10^{6}$ & 10.25 & 1.55 & 0.125 &  0.006\\
              \hline
              $J_P$  ($10^{-10}$ photons/s) & 70 & 1.32 & 0.002 & 1.85 10$^{-5}$ \\
              \hline
              $J_D $ (photons/s) & 30 & 19.4 & 12 & 1.25\\
              \hline
              $\Delta J_{\mathrm{PHE}}$  ($10^{-5}$ photons/s) & 30 & 2.9 & 0.15 & 0.0008 \\
              \hline
              $a_{\mathrm{PHE}}$ ($\mu$m/s$^2$) & 2.8 & 0.65 & 0.0085 & 5 10$^{-5}$ \\
              \hline
 \end{tabular}
      \caption{Relevant numerical values for different atomic elements. The estimate of the quantities $\sigma_{\mathrm{PHE}}$ (the PHE scattering cross-section), the ratio of spontaneous emission rate and natural decay rate $A_D/\Gamma_D$ (without Doppler broadening, the maximum value of $1$ occurs in the absence of other inelastic decay channels), circular frequency detuning $\Delta_{PD}$ between $S$ and $P$-level, the currents $J_P =\rho_P A_P$, $J_Q=\rho_Q A_Q$ generated by ED and EQ scattering, and  the photonic Hall current $\Delta J_{\mathrm{PHE}}$ which equals the difference between up and down current along $y$ axis (all expressed in number of photons per second), $a_{\mathrm{PHE}}$ (magneto-transverse recoil acceleration). The PHE is normalized by the magnetic quantum number $m_D= \pm 1, \pm 2$ of the $D$-level that is resonantly excited.  The electronic spin and the Zeeman structure of the ground state, for the ions, are disregarded in the calculation of the various quantities.}
      \label{table}
\end{table}

In this section, we derive the photon Hall scattering cross-section and current, considering a resonant EQ transition and an off-resonant ED transition. The calculation is performed for one single atom at rest in interaction with a {quasi-}monochromatic light field. More realistic situations are discussed in the following section.

Our starting point is the Kramers-Heisenberg formula for the optical cross-section \cite{loudon2000quantum}:
\begin{eqnarray}\label{eq_totaldiffcross}
    \frac{d\sigma}{d\Omega}(&\mathbf{k}&\varepsilon\rightarrow\mathbf{k}_s\varepsilon_s)= \nonumber \\ &\alpha^2&\frac{\omega_s^3}{\omega^3}|f_{ED}(\omega,\varepsilon,\varepsilon_s)+f_{EQ}(\mathbf{k},\mathbf{k_s}\varepsilon,\varepsilon_s)|^2.
\end{eqnarray}
Here, $\alpha$ is the fine structure constant; $\mathbf{k}$ and $\mathbf{k_s}$ are the wave vectors respectively for the incident and the scattered field;  $\omega_{(s)}=|\mathbf{k_{(s)}}|c_0$ is the frequency of the radiation and $c_0$ the speed of light in vacuum; $\varepsilon$ and $\varepsilon_s$ are the incident and the scattered field polarization; $f$ is the complex scattered field amplitude either for the ED transition or for the EQ transition. This model is valid in the weak field intensity limit where the population of all levels is far from saturation, i.e. a level population $\rho \ll \frac{1}{2}$. For a random incident polarization, the population is given by \cite{loudon2000quantum}

\begin{eqnarray}\label{popu}
    \rho= \frac{B}{2\pi}\int d \Delta \frac{W(\omega_0 + \Delta)}{\Delta^2 + \gamma^2}
\end{eqnarray}
 In this formula, $\Delta$ is the detuning from the resonance ($\omega_0=\omega_P$ for the ED transition, $\omega_0=\omega_D$ for the EQ transition), $B= A/W_0$ ($W_0=8\pi \hbar/\lambda^3$)  is the stimulated emission coefficient, $A$ the natural linewidth, and $\gamma$ the total linewidth ($\Gamma \equiv 2 \gamma$ is the Lorentzian FWHM linewidth), and $W$ is the radiation density per (circular) frequency interval.
In the following we will focuss on one magnetic sub level of the $D$-level that is  typically separated energetically by an amount $\mu B$ from the others due to an applied bias magnetic field. We shall assume that $\gamma_L \ll \Gamma_D \ll \mu B /\hbar$, i.e. that the laser bandwidth $\gamma_L$ is small enough to resolve the magnetic sub levels and is put on resonance with one of them, and the Zeeman shift $\mu B/\hbar$ lifts the magnetic degeneracy completely. Moderate magnetic fields of about $10$\,G are enough to satisfy the above inequality. The linewidth $\Gamma_D$ may exceed the natural linewidth $A_D$, due to inelastic scattering to other levels. Doppler broadening will be discussed in the next section. If the above inequalities are satisfied,  we obtain $\rho_D= (B/2\pi \gamma_D^2)\times (I/c_0)\equiv \frac{1}{2}I/I_s$, with $I=W \gamma_Lc_0$ the flux density and $I_s$ the saturation intensity given by,
 \begin{eqnarray}\label{sat}
    I_s= \frac{\pi hc_0 \Gamma_D^2}{A_D\lambda_D^3}
\end{eqnarray}

For the population of the $P$-level, Eq.~(\ref{popu}) reduces to  $\rho_P = (B_P/2\pi) W\gamma_L/\Delta^2 = (I \lambda_P^3 /8\pi hc_0) A_P/\Delta^2$. Because of the far off-resonant excitation (see values of $\rho_P$ in Table \ref{table}), the ED scattering is completely negligible.

The differential cross-section for elastic photon Hall scattering, given random incident polarization, involves the interference of the ED transition (ED moment $r_P$) and the EQ transition (EQ moment $q_D$) \cite{PHEatom},
\begin{eqnarray}\label{diffcross}
    \frac{d\sigma}{d\Omega}(\mathbf{k},\mathbf{k}_s,\mathbf{B}, m_D) = 2\alpha^2\mathrm{Re} \left[f_{ED}f_{EQ}\right] = \frac{\alpha^2\omega^6}{c_0^4}r_P^2q_D^2\nonumber \\
    \times \mathrm{Re} \left[ i\mathbf{k}_S\cdot \hat{\mathbf{y}}\sum_m F_{mm_D}A_{mm_D}(\mathbf{k},\mathbf{k}_s,\mathbf{B}) \right]
\end{eqnarray}
where $\hat{\mathbf{y}}$ is the magneto-transverse unit vector $\hat{\mathbf{y}}=\hat{\mathbf{k}}\times \hat{\mathbf{B}}$. For a strongly detuned $P$-level and a laser tuned resonantly ($\Delta_D=0, \gamma_L\ll \Gamma_D$) on one Zeeman substate $m_D$ of the $D$-level, it follows that the PHE scattering amplitude is given by
\begin{equation}\label{pheamp}
    F_{mm_D}= -\frac{2i}{\Gamma_D \Delta_{DP}}.
\end{equation}

$\Gamma_D$ is the total decay rate of the $D$ state and $\Delta_{PD}$ the frequency difference between the $P$ and $D$ state. Since we have to sum over all Zeeman substates of the $P$-level, the angular functions add up to
\begin{eqnarray*}
    \sum_{m=0,\pm 1} A_{mm_D} = \frac{m_D}{60}\left[1-2(\mathbf{k}_s\cdot \hat{\mathbf{z}})^2 \right] \delta_{m_D=\pm 1} \nonumber \\
+ \frac{m_D}{60}\left[1+(\mathbf{k}_s\cdot \hat{\mathbf{x}})^2 -\frac{1}{2}(\mathbf{k}_s\cdot \hat{\mathbf{y}})^2 \right] \delta_{m_D=\pm 2}
\end{eqnarray*}
where $\hat{\mathbf{x}}$ is directed along the incident wave vector $\mathbf{k}$ and $\hat{\mathbf{z}}$ along the magnetic field $\mathbf{B}$. The total photon Hall flux directed along $\hat{\mathbf{y}}$ is $\Delta J_{\mathrm{PHE}}= \sigma_{\mathrm{PHE}} I_{\mathrm{in}}$, with
\begin{eqnarray}
  \sigma_{\mathrm{PHE}} &=&  \frac{\pi m_D}{75}\frac{\alpha^2\omega^6}{\Delta \Gamma_Dc_0^4}r_P^2q_D^2 \left[\delta_{m_D=\pm 1} + \frac{3}{2}\delta_{m_D=\pm 2}\right]
\end{eqnarray}

It is convenient express this in terms of the natural linewidth. For the two transitions is
\begin{equation}\nonumber
    A_D= \frac{e^2k_D^5q_D^2}{300\pi \varepsilon_0 \hbar}
\end{equation}
and
\begin{equation}\nonumber
    A_P= \frac{e^2k_P^3r_P^2}{3\pi \varepsilon_0 \hbar}
\end{equation}
Hence
\begin{eqnarray}
  \sigma_{\mathrm{PHE}} &=& m_D\frac{3 \lambda_D^2 }{32\pi} \frac{\omega_D^3}{\omega_P^3} \frac{A_PA_D}{\Delta_{PD}\Gamma_D}
  \left[\delta_{m_D=\pm 1} +\frac{3}{2}\delta_{m_D=\pm 2}\right]
  \label{eq_sigma_PHE}
\end{eqnarray}

Expression~(\ref{eq_sigma_PHE}) is the most important result of this paper. It shows that a PHE current exists along the $\hat{\mathbf{y}}$ direction.  As long as the bias magnetic field lifts the degeneracy of the Zeeman substates of the $D$-level, and is resonantly tuned on one such substate, the PHE current is just proportional to the magnetic quantum number $m_D$. This result is different from the situation studied in Ref.~\cite{PHEatom} that assumes all magnetic sublevels, both of $P$ and of $D$-levels, to be resolved by a broadband beam ($\gamma_L >> \mu B, \Delta$). In that case the PHE is proportional to the magnetic field. Compared to Ref.~\cite{PHEatom}, the PHE cross-section is suppressed because of the large detuning of the $P$-level ( $A_P/\Delta_{PD} \ll 1)$.
On the other hand,  by the absence ED scattering, the background noise of normally scattered photons is small as well. We can estimate the optical cross-section at wavelength $\lambda_D$ for resonant EQ-scattering as $\sigma_{\mathrm{EQ}}=\rho_D A_D/I \times \hbar \omega_D$. This yields, using Eq.~(\ref{popu}),
\begin{eqnarray}
  \sigma_{\mathrm{EQ}} &=& \frac{\lambda_D^2}{2\pi}  \left(\frac{A_D}{\Gamma_D}\right)^2
  \label{eq_sigma_EQ}
\end{eqnarray}
If the photon scattering from the EQ transition can be adopted as the main source of background noise, the signal-to-noise ratio is $\sigma_{\mathrm{PHE}}/\sigma_{\mathrm{EQ}}\propto (A_P/\Delta_{PD})\times (\Gamma_D/A_D) $. This would benefit from a very low EQ branching ratio, and thus makes neutral Strontium a favorite candidate.
From Table \ref{table} we can infer that for neutral Strontium, at $I=I_s$, $J_{EQ} \approx 30/2 =15 $ photons/second are scattered by the electric quadrupole along the $y$-direction. This is much larger than the scattered rate of thermal photons we expect at room temperature $\sim\Gamma_D\,\textrm{exp}(-hc/(\lambda_Dk_BT))\sim 4\cdot10^{-11}\,$s$^{-1}$. The PHE generates  $3 \cdot 10^{-4}$ photons per second. Note that the order of magnitude of the relative Hall effect $ \sigma_{\mathrm{PHE}}/\sigma_{\mathrm{EQ}}  $ is of order $10^{-6} - 10^{-8}$, quite similar to the ratios reported for classical samples \cite{geert}.  The PHE current gives rise to a magneto-transverse recoil force, $ F_{\mathrm{PHE}}= Ma_{\mathrm{PHE}}= \Delta J_{\mathrm{PHE}} /c_0$, with $M$ the atomic mass.

\section{Implementation and Doppler Effect}
\label{Implementation}

For practical implementation, one might consider a large ensemble of cold atoms ($N \sim 10^8$), hold together in a far off-resonance dipole trap with, at least, a confinement along the $z$-axis to prevent atoms falling due to the Earth gravitational attraction. One expects a large Doppler broadening of the EQ-transition. Indeed, even with efficient laser cooling, the atomic velocity distribution gives rise to a Gaussian profile with a frequency linewidth of the order of $\Gamma_v/2\pi\sim 10^5$\,Hz, much larger than $\Gamma_D$.

We assume to be in the regime $\Gamma_D \ll \Gamma_v , \gamma_L\ll \mu B/\hbar$, where $\gamma_L$ is the laser bandwidth. The laser is still assumed to be tuned on the EQ resonance, but its bandwidth is sufficiently broadened to make sure to address the full Doppler distribution of atoms. This broadening can be  achieved using an external frequency modulation device such as an acousto-optic modulator. Because the natural linewidth $\gamma$ (decay rate $\Gamma = 2 \gamma$) is much smaller than both the laser bandwidth and the Doppler width, we can write for the Doppler shifted line profile of one atom with velocity $v$, $1/[(\Delta-\omega_0v/c_0)^2 + \gamma^2]
\approx (\pi/\gamma) \delta(\Delta - \omega_0v/c_0)$. As a result, when averaged over the whole atoms distribution, Eq.~(\ref{popu}), applied to the state $D$, modifies to,

\begin{eqnarray}\label{popudopp}
    \rho_D &=& \frac{B_D /2 \gamma_D}{\sqrt{2\pi }v_T} \int_{-\infty}^{\infty} dv \exp\left(\frac{-v^2}{2v_T^2}\right) {W(\omega_0+\omega_0v/c_0)} \nonumber \\
    &\approx & \frac{A_D}{\Gamma_D}\frac{W}{W_0}\textrm{erf}\left(\frac{\gamma_Lc_0}{2\sqrt{2} \omega_0 v_T}\right) \nonumber\\
    &= & \frac{A_D}{\Gamma_D}\frac{I \lambda_D^3}{4h\gamma_L c_0}\textrm{erf}\left(\frac{\sqrt{\log 2}}{2}\frac{\gamma_L}{\Gamma_v}\right)
\end{eqnarray}
where $v_T= \sqrt{k_BT/M}$ is the typical thermal velocity of the atoms along the incident wave vector; $\textrm{erf}(x)$ is the error function, which arises here because  we adopt a flat laser intensity $W= I/\gamma_Lc_0$ between $\omega_0 - \gamma_L/2$ and $\omega_0 + \gamma_L/2$.  For a  laser bandwidth $\gamma_L= 4\Gamma_v$, $\textrm{erf}(x)\simeq 0.98$, we find that a laser intensity $I_v=\eta I_s$, with $\eta\approx 2 \Gamma_L/(\pi\Gamma_D)$ induces a population $\rho_D = 1/2$, \emph{i.e.} without saturating the $D$-level. For $\Gamma_D/2\pi \approx 500$ Hz (neutral Strontium) and
$\gamma_L/2\pi = 4 \times \Gamma_V /2\pi \approx 4\cdot 10^5$ Hz, we estimate $\eta \approx 500$, leading to maximal intensities of order of $100$ W/m$^2$ (Table 2). The incident laser power should thus be increased by this factor $\eta$. Such a laser power is reachable with the current laser technology. We note that the population of the $P$ state also increases  by the  factor $\eta$ but still remains negligible.
Similarly, it is straightforward to see that that the Doppler broadening reduces the PHE complex amplitude (\ref{pheamp}) by the same factor
\begin{equation}\label{pheampdopp}
    F_{mm_D}= -\frac{1}{\eta}\frac{2i}{\Gamma_D \Delta_{PD}}.
\end{equation}
This means that the PHE cross-section $\sigma_{\mathrm{PHE}}$ also becomes a factor $\eta  $ smaller. The same holds for $\sigma_{EQ}$. This means that the currents $\sigma_{\mathrm{PHE}}\times I_v$ and $\sigma_{\mathrm{EQ}}\times I_v$ both remain constant. We conclude that, provided the intensity is increased by a factor $\eta$ over a bandwidth a few times larger than the Doppler broadening, the estimates for $\Delta J_{\mathrm{PHE}}, J_D, a_\mathrm{PHE}$ and $\sigma_\mathrm{PHE}/\sigma_\mathrm{EQ}$ in Table 1 will be unaffected by the Doppler effect (see Table II for a summary of the typical values). For a cloud with  $N= 10^8$ atoms the PHE involves roughly  $3 \cdot 10^4$ photons per second.

The photonic Hall current gives rise to a photonic Hall force, $F= Ma= 2 J_{\mathrm{PHE}} /c_0$, thus a net displacement $y_{\mathrm{PHE}}$ of the atomic cloud centre of mass.  A much stronger displacement occurs along $\hat{\mathbf{k}}$ due to resonant absorption of the incident laser beam, which rapidly pushes the atoms out of resonance. To remove this spurious effect, one can periodically reverse the direction of the incident beam. At the same time, the direction of the magnetic field must be reversed as well to make sure that the PHE force is always pointing in the same direction.  A direct measurement of the displacement of the cloud of atoms due to the PHE is then mainly limited by the heating rate associated with the elastic and inelastic photon scattering. If we assume a cycle of an absorption and re-emission of a photon with subsequent opposite directions at each life time $1/\Gamma_D$ of the $D$-level, the velocity dispersion along the $y$-axis will grow in time as $\Delta v_y^2\simeq v_r^2\Gamma_Dt/3$, where $v_r=\hbar k_D/M$ is the recoil velocity. Here the scattering rate can be  overestimated by at least a factor of two for a close  $S\leftrightarrow D$ transition and should depend on the decay channels toward the $S$ state in the  presence of inelastic scattering. The spatial confinement along the $y$-axis  then grows as $\Delta y^2\simeq \Delta v_y^2t^2 \simeq v_r^2\Gamma_Dt^3/3$. The magneto-transverse displacement is estimated as $y_{\mathrm{PHE}} = \frac{1}{2}a_{\mathrm{PHE}} t^2= \Delta J_{\mathrm{PHE}}v_r t^2$ in terms of the difference $\Delta J_{\mathrm{PHE}}$ of the number of photons per second along negative and positive $y$-axis. In Table \ref{table2}, we estimate the ratio $y_{\mathrm{PHE}}/\sqrt{\Delta y^2} \approx
 \Delta J_{\mathrm{PHE}} \sqrt{t/\Gamma_D}$ after $t=1$ second. The most promising value of around  $5\cdot 10^{-6}$ is obtained for neutral Strontium ($y_{\mathrm{PHE}}\sim 1 \, \mu$m). This  small value should be observable with a large number of atoms ($N=10^8 - 10^9$) \cite{yang2015high}.

\begin{table}[h]
\begin{tabular}{| c | c | c | c | c |}
                \hline
               Element & Sr & Ca & Yb$^+$ & Sr$^+$ \\
               \hline\hline
               ED line & \multicolumn{2}{|c|}{$^1S_0\leftrightarrow\,^1P_1$} & \multicolumn{2}{|c|}{$^2S_{1/2}\leftrightarrow\,^2P_{3/2}$} \\
               \hline
              $\lambda_P$ (nm) & 461  & 423 & 329  & 408 \\
              \hline
              $A_P/2\pi$ (MHz) & 32 & 34 & 19.6 & 22 \\
              \hline\hline
              EQ line & \multicolumn{2}{|c|}{$^1S_0\leftrightarrow\,^1D_2$} & \multicolumn{2}{|c|}{$^2S_{1/2}\leftrightarrow\,^2D_{5/2}$}\\
              \hline
              $\lambda_D$ (nm) & 496 & 458 & 411 & 674 \\
              \hline
              $A_D/2\pi$ (Hz) & 9.5 & 6.2 & 3.8 & 0.4 \\
              \hline\hline
              $\Gamma_v/\Gamma_D$ & 211  & 1935 &  4473 & 2.5 $10^5$\\
              \hline
              $I_v$ (W/m$^2$) & 113 & 22 & 21 & 0.8\\
              \hline
              $\Delta J_{\mathrm{PHE}}$  ($10^{-5}$ photons/s) & 30 & 2.9 & 0.15 & 0.0008 \\
              \hline
              $a_{\mathrm{PHE}}$ ($\mu$m/s$^2$) & 2.8 & 0.65 & 0.0085 & 5 10$^{-5}$ \\
              \hline
              $y_{\mathrm{PHE}}/\sqrt{\Delta y^2} \times 10^6$ (@ 1s) & 5.5& 1.6 & 0.13 & 0.005\\
              \hline
 \end{tabular}
      \caption{The same information as in Table 1 but this time assuming a Doppler  broadening of $\Gamma_v/2\pi= 10^5$ Hz.  The photonic Hall current $J_{\mathrm{PHE}}$ and the magneto-transverse recoil acceleration $a_{\mathrm{PHE}}$ are not changed provided the measurement is done at the same $D$-level population. The bottom line roughly estimates the mean magneto-transverse displacement of atoms after one second, normalized to the typical size of the expanding cloud due to EQ absorption of photons, $y_{\mathrm{PHE}}/\sqrt{\Delta y^2} \sim \sqrt{t}$ .}
      \label{table2}
\end{table}

Direct observation of a cloud displacement by PHE might be not suitable for an ion cloud. First because ion clouds are usually strongly confined in a Paul trap for example. Secondly, the atom number is usually lower than the equivalent experiment with neutral atoms.
For ions, we suggest to measure the scattering imbalance due to PHE, that is the equivalence of scattered photons along positive and negative  $y$-axis.  This method can actually also be applied to neutral atoms system and was already employed to observe the PHE in classical samples \cite{geert}, which exhibited the same orders of magnitude. It offers several advantages with respect to the observation of a cloud displacement. First, one can use an optical frequency filter to get ride of the unwanted inelastic scattering. Second, a synchronous modulation technique can be employed to remove large noise contributions, if synchronized  with the fluctuating direction of the incident field to keep the atoms at rest.  Indeed, by reserving periodically the direction of the incident beam, the PHE appears with the same period whereas the fluorescence background is not observed  because it is independent of the sign of the wave vector. Residual systematic errors should be subtracted repeating the experiment with a reverse magnetic field.

\section{Conclusions}
We have shown that the PHE occurs in atomic or ionic species having an electric-dipole and an electric-quadrupole transition that are far but not too far separated in energy.
The excitation optical field is tuned on resonance with a specific magnetic sublevel of the $D$-level. In that case the PHE does not depend on the strength of the magnetic field used to lift the Zeeman degeneracy of the $D$ state. Since the $P$-transition is off-resonant the $S\leftrightarrow P$ transition scatters no photons, and the PHE only relies on the dispersive response of the electric-dipole transition.

We suggest two methods to observe PHE. One method is based on the observation of the cloud displacement induced by the PHE force. The second method consists of  measuring the elastic photon scattering imbalance. In all practical cases, discussed in this paper, the PHE leads to a weak signal, mainly because of the large frequency detuning of the $P$. Nevertheless, the effect should be observable with the current state-of-the-art experimental platforms. A more spectacular PHE can be expected if one can manage to increase the electric-multipole transition strengths with respect to the electric dipole transition. One promising road consists of placing atoms close to a two-dimensional plasmonic material, where the field confinement, induced by a plasmonic excitation, strongly reduces its effective wavelength, as was recently suggested in \cite{rivera2016shrinking}.

\bibliographystyle{apsrev}

\end{document}